\documentclass[prd,showpacs,preprintnumbers,nofootinbib,aps]{revtex4}
\usepackage{subeqnarray}
\usepackage{epsfig,amsmath,amssymb}
\usepackage{mathrsfs}
\usepackage[usenames,dvipsnames]{color}
\usepackage[pagebackref=false, colorlinks=false]{hyperref}
\definecolor{redish}{rgb}{0.7,0.2,0.0}  
\definecolor{bluish}{rgb}{0.2,0.5,0.8}
\hypersetup{linkcolor=redish,          
                  citecolor=blue,        
                  filecolor=magenta,      
                  urlcolor=bluish}          
\DeclareFontFamily{U}{rsfs}{}         
\DeclareFontShape{U}{rsfs}{m}{n}{<5> rsfs5 <6><7> rsfs7          %
  <8><9><10><10.95><12><14.4><17.28><20.74><24.88> rsfs10}{}     %
\DeclareMathAlphabet{\mathfs}{U}{rsfs}{m}{n}                     %

\newcommand{\ba}{\nopagebreak[3]\begin{eqnarray}}
\newcommand{\ea}{\end{eqnarray}}
\newcommand{\bii}{\begin{itemize}}
\newcommand{\eii}{\end{itemize}}

\begin{document}

\title{Tachyon Perturbation on Two Dimensional Black Hole }
\author{Aniket Basu}
\email{aniket.basu@gmail.com}
\affiliation{Vidyasagar College, Kolkata 700006, India}
\author{Parthasarathi Majumdar}
\email{parthasarathi.majumdar@rkmvu.ac.in}
\affiliation{Ramakrishna Mission Vivekananda University, Belur Math 711202, India}

\begin{abstract}
We study the geometry of the two dimensional string theoretic black hole under tachyonic perturbations. These perturbations are restricted to affect only the metric and the dilaton, while other string theoretic excitations (like the axion) are ignored. The metric and linearized dilaton perturbations are determined to lowest non-trivial order of the tachyonic hair in the presence of back reaction. We evaluate the Kretschmann scalar and argue that the horizon does not become singular in the presence of tachyon perturbations (to the order of our consideration). A closed-form solution of the allowed tachyon field and that of the allowed tachyon potential emerges as a requirement of self-consistency of our solution.
\end{abstract}

\maketitle

\section{Introduction}

The possibility that a black hole is actually a low energy approximation of an exact two dimensional conformal field theory, raised towards the end of the last century \cite{Witten:1991yr}, evoked an explosion of interest in the search for black hole solutions in 1+1 spacetime dimensions \cite{Mandal:1991tz, Rocek:1991vk}. The early black hole solutions ignore most string excitations except the metric and the dilaton field \cite{Mandal:1991tz}, and the task of incorporating other massless stringy backgrounds like the tachyon and the axion fields is left in this body of work to posterity. Whether this two dimensional spacetime retains its geometry when the tachyon field is turned on, thus remains an important issue.  While the subsequent literature on this topic does indeed include the effect of tachyons \cite{DeAlwis:1991vz, KalyanaRama:1993np, KalyanaRama:1993hk}, apparently a proper assessment of the effects of a tachyonic field on the black hole geometry has not quite been made \cite{Peet:1993vb}. Thus, claims that the horizon may actually become singular in presence of tachyons \cite{KalyanaRama:1993np, KalyanaRama:1993hk} seemingly warrant a revisit for arriving at unambiguous conclusions. More recently, the incorporation of axion fields have been argued to destabilize the horizon to the extent that it may not form \cite{Karczmarek:2004bw}. These developments have been reviewed in \cite{Grumiller:2002nm, Grumiller:2006rc}.   

In the present paper, we focus on the simpler issue of assessing effects of perturbative inclusion of a tachyonic field to the system considered in the original action \cite{Mandal:1991tz}. Other massless string excitations like the axion are ignored here. Further, our current assay is restricted to the assumption that the effect of turning on the tachyonic field is to perturb (linearly) the dilaton and the metric,  and we attempt to compute these perturbations as well as the form of the tachyonic hair as a result, and therefore attempt to derive self-consistently the allowed form of a non-vanishing tachyon potential. 
 
The rest of this paper is organized thus: In section \ref{review}, we review the black hole solution of \cite{Mandal:1991tz}. In subsequent sections, working in the Penrose conformal frame, we evaluate the perturbed metric as follows: In section \ref{perturbation}, we consider the effect of turning on the tachyon, and derive the relevant equations (differential and algebraic) for the metric perturbation and the dilaton perturbation,  to lowest non-trivial order in the tachyon field. In section \ref{solution}, we solve these equations for the tachyon, the metric perturbation and the dilaton perturbation in the vicinity of the horizon.  In section \ref{OT^2},  we consider the effect of including $O(T^2)$ source terms in our equations of motion, and solve for the metric and linearized dilaton perturbations. In section \ref{nearhorizonbehaviour}, we consider the near horizon behaviour of the perturbed metric, and evaluate the Ricci scalar and the Kretschmann scalar to show that in the neighbourhood of the horizon, these quantities remain finite everywhere and everywhen in the presence of a tachyonic perturbation. Calculational details involving the solutions of the perturbed metric and linearized dilaton fields  have been included in Appendices \ref{W} and \ref{Phi_1}.

\section{A Brief Look at the Original Solution}
\label{review}

In this section we briefly review the black hole solution of \cite{Mandal:1991tz}, in which the authors considered the target space action
\begin{equation}
S=\int d^2 x \exp(-2\Phi)\sqrt{G}\left[R-4(\nabla\Phi)^2+(\nabla T)^2+V(T)\right]
\end{equation}
where 
\begin{equation}
V(T)=-\frac{2}{\alpha'}T^2+O(T^3)
\end{equation}
is the tachyon potential. The equations of motion are then:
\begin{eqnarray}
R_{\mu\nu}-2\nabla_\mu\nabla_\nu\Phi+\nabla_\mu T\nabla_\nu T=0\\
R+4(\nabla\Phi)^2-4\nabla^2\Phi+(\nabla T)^2+V(T)+c=0\\
-2\nabla^2 T+4\nabla\Phi\nabla T+V'(T)=0
\end{eqnarray}
where $c=(D-26)/(3\alpha')=-8/\alpha'$ as $D=2$ in this case. To solve this set of coupled equations for $G_{\mu\nu}, \Phi$ and $T$ is quite difficult. The incipient approach is to solve it for the case $T=0$, when the above set of equations reduce to 
\begin{eqnarray}
R_{\mu\nu}-2\nabla_\mu\nabla_\nu\Phi=0\\
R+4(\nabla\Phi)^2-4\nabla^2\Phi+c=0
\end{eqnarray}

\subsection{Schwarzschild Frame}
Choosing a gauge where the dilaton is proportional to one of the coordinates, $\Phi=Q\eta/2$, as in  \cite{Mandal:1991tz}, a one-parameter solution emerges:
\begin{equation}
ds^2=-g(\eta)dt^2+\frac{1}{g(\eta)}d\eta^2
\end{equation}
where $g(\eta)=1-M \exp(Q\eta)$ with $Q^2=8/\alpha'$. Here $M$ is a `mass'-like parameter characterizing the black hole, which has a horizon at $\eta=-\frac{1}{Q}\ln M$. The scalar curvature is given by
\begin{equation}
R(\eta)=-MQ^2 \exp(Q\eta)
\end{equation}
which is clearly finite at the horizon.

\subsection{Conformal Frame}
In  \cite{Mandal:1991tz} we also find a solution in the conformal gauge, defined as
\begin{eqnarray}
ds^2=e^\sigma(dx^2-dy^2)=e^\sigma du dv
\end{eqnarray}
where the light-cone coordinates $u=x+y, v=x-y$ are related to $\eta, t$ by a series of coordinate transformations which we will outline in the next subsection.

The Ricci tensor has components:
\begin{eqnarray}
R_{uu}=R_{vv}=0, R_{uv}=\partial_u\partial_v\sigma
\end{eqnarray}
and the Ricci scalar is 
\begin{equation}
R=4e^{-\sigma}\partial_u\partial_v\sigma
\end{equation}

Then the equations of motion reduce to 
\begin{eqnarray}
R_{uv}=2\partial_u\partial_v\Phi\\
R_{uu}=2\nabla_u\partial_u\Phi\\
R_{vv}=2\nabla_v\partial_v\Phi
\end{eqnarray}

Solving the first leads to $\sigma=2\Phi$. Solving the rest leads to
\begin{equation}
ds^2=\frac{dudv}{2uv/\alpha'+M}
\end{equation}
which is a two dimensional analogue of the Schwarzschild solution in Kruskal-Szekeres coordinates. The horizon, according to \cite{Mandal:1991tz}, is given by the lines $uv=0$ and the curvature may be shown to be
\begin{equation}
R=\frac{-8M/\alpha'}{2uv/\alpha'+M}
\end{equation}
which does not blow up at the horizon. The singularity occurs at $uv=-M\alpha'/2$. (We note in passing that we subsequently define $c_M=M\alpha'/2$, and redefine our coordinates such that $\bar{u}\bar{v}=uv/c_M$, and this we expect to satisfy $\bar{u}\bar{v}=-1$ at the singularity, and indeed it turns out it does.)

\subsection{Coordinate transformation linking these two frames}
The relation between $(\eta, t)$ and $(u,v)$ is given by
\begin{eqnarray}
u=\exp\left[-\frac{Q}{2}(\eta^*+t)\right]\\
v=-\epsilon\exp\left[-\frac{Q}{2}(\eta^*-t)\right]\\
\frac{\alpha'}{2}\exp(Q\eta^*)=\frac{\exp(Q\eta)}{1-M\exp(Q\eta)}
\end{eqnarray}
where $\epsilon=+1$ `inside' the horizon and $\epsilon=-1$ `outside'.


\section{Turning on the tachyon}
\label{perturbation}

With minor modifications, we follow the notation of \cite{Mandal:1991tz}, in which the black hole solution in the conformal gauge was given in terms of $\sigma$ and $\Phi$ in the case where the tachyon field $T$ is set to $0$. Now if we introduce a non-trivial tachyonic field $T$, clearly the effect is to perturb $\sigma$ and $\Phi$:
\begin{eqnarray}
\nonumber \sigma=\sigma_0+\sigma_1\\
\nonumber \Phi=\Phi_0+\Phi_1
\end{eqnarray}
where $\sigma_0$ and $\Phi_0$ are the values of these fields when $T=0$, and $\sigma_1$ and $\Phi_1$ are possible corrections that arise due to the introduction of the tachyon. Note that these corrections are perturbative in the tachyon field to {\it lowest non-trivial order}. In what follows, we try to determine whether the Ricci scalar remains finite everywhere in spacetime as a result of the perturbation. Our approach entails certain assumptions which we specify below; our objective is to ascertain what transpires upon incorporation of the tachyon, for the metric and the Ricci scalar. In the absence of the tachyon, the Ricci scalar is
\begin{eqnarray}
R=R_0=-\frac{8M}{\alpha'}e^{\sigma_0}
\end{eqnarray}
which does not become singular at the horizon. Does this situation continue to hold, when tachyon perturbations are included?

It can be shown that our set of partial differential equations in $\sigma$ and $\Phi$ (in the conformal frame) reduce to 
\begin{eqnarray}
\partial_u\partial_vW=-\partial_u T\partial_v T \label{sig1}\\
\partial^2_uW-\partial_u\sigma_0\partial_uW=-(\partial_uT)^2 \label{sig2}\\
\partial^2_vW-\partial_u\sigma_0\partial_vW=-(\partial_vT)^2 \label{sig3}\\
\partial_u\partial_v[W-2\Phi]+4\partial_u\Phi\partial_v\Phi+\partial_uT\partial_vT+\frac{e^\sigma}{4}V(T)-\frac{2e^\sigma}{\alpha'}=0 \label{Wphi}
\end{eqnarray}
where $W \equiv \sigma_1-2\Phi_1$.

In addition, we have an equation of motion for the tachyon, which happens to be
\begin{equation}
-2\nabla^2 T+4\nabla\Phi\nabla T+V'(T)=0 \label{tacheq}
\end{equation}
which we shall have to solve for the allowed form of $T$.

To solve the first three of eqn.s (\ref{sig1})-(\ref{tacheq}), as a first approximation, we keep only terms linear in $T$ and its derivatives, yielding
\begin{eqnarray}
W_{uv}=0\\
W_{uu}-\sigma_{0u}W_u=0\\
W_{vv}-\sigma_{0v}W_v=0
\end{eqnarray}
which can be immediately solved to obtain the solution $W=const.$ Since our equations involve derivatives of $W$, it is obvious that  the actual value of this constant will not matter. Thus, to this order of the tachyonic perturbation, the consequent metric and dilaton perturbations are essentially identical, modulo an additive constant. Note that, to arrive at this solution, we have used the fact that \cite{Mandal:1991tz}
\begin{equation}
\sigma_0=2\Phi_0=-\ln\left[\frac{2uv}{\alpha'}+M\right]
\end{equation}
where $M$ is a {\it dimensionless} ``mass'' parameter for the black hole.

\section{Solving for the metric perturbation $\sigma_1$ and the tachyon $T$}
\label{solution}

\subsection{Solving for the metric perturbation $\sigma_1$}

Thus, to this order of the tachyonic perturbation, eqn. (\ref{Wphi}) for the metric perturbation reduces to 
\begin{equation}
-\partial_u\partial_v\sigma_1+\sigma_{0u}\sigma_{0v}+\sigma_{0u}\sigma_{1v}+\sigma_{1u}\sigma_{0v}-\frac{2e^\sigma}{\alpha'}=0
\end{equation}

It is not simple to solve the full equation,  so we give approximate solutions for the regions $uv \gg c_M$ and $uv \ll c_M$; in these regions one obtains the equations 
\begin{eqnarray}
uv\sigma_{1uv}+(u\sigma_{1u}+v\sigma_{1v}+e^{\sigma_1})-1 & = & 0~,~uv \gg c_M \label{gg} \\ 
c_M\sigma_{1uv}+(u\sigma_{1u}+v\sigma_{1v}+e^{\sigma_1}) &= & 0~,~uv \ll c_M~ \label{ll}
\end{eqnarray}

Eqn (\ref{gg}) can be approximately solved in the asymptotic region, leading to the solution $\sigma_1 = const. (uv)^{-1}$. 

Eqn (\ref{ll}) has the solution 
\begin{equation}
\sigma_1=-uv/c_M+\ln{B}
\end{equation} where $B$ is a constant, and, consequently
\begin{equation}
\Phi_1=-uv/2c_M+\mathrm{constant} ~,
\end{equation}
where, $c_M \equiv M (\alpha'/2)$.

This has the consequence that, to this order in the tachyon perturbation, the horizon undergoes a shift in the $uv$ plane, leading to the idea of a `perturbed' horizon. However, there is apparently no singularity on the perturbed horizon.  We justify this last statement in section \ref{nearhorizonbehaviour}.

In Section \ref{OT^2} we relax the restriction that the metric perturbations are only linear in the the tachyon field and its derivatives, and proceed to consider the more general case.

\subsection{Solving for $T$}
\label{T}

The equation of motion for $T$ becomes
\begin{equation}
4(uv+c_M)T_{uv}+2(uT_u+vT_v)+T=0
\end{equation}
where $c_M=M\alpha'/2$. 

Here we depart from the assumption of the form of the tachyon solution adopted in \cite{Witten:1991yr, Mandal:1991tz}: these authors adopted the ansatz $T=\tilde{T}\exp(\Phi)$, but then solving for $\tilde{T}$ as a function of $u,v$ does not seem simple. 

Instead, we try a different track. Assuming separability of $T$ as a product of functions of $u,v$, $T(u,v)=U(u)V(v)$, it may be shown that
\begin{equation}
T=\frac{\mathrm{constant}}{\sqrt{(u-\kappa\sqrt{c_M})(v+\sqrt{c_M}/\kappa)}}
\end{equation}
where $\kappa$ is a non-zero constant independent of $u,v$ and originates from the equation
\begin{eqnarray}
\frac{1+2u\frac{U'}{U}}{2\sqrt{c_M}\frac{U'}{U}}=-\frac{2\sqrt{c_M}\frac{V'}{V}}{1+2v\frac{V'}{V}}=\kappa
\end{eqnarray}

This can become large at $u=\kappa\sqrt{c_M}$ and $v=-\sqrt{c_M}/\kappa$. When $T$ is large then the equations for $W$ and $T$ used here are no longer valid as the nonlinear terms become more important. (Note that this expression for $T$  blows up, as expected, at the singularity $uv=-c_M$, but perhaps the expression becomes meaningless much before the singularity has been reached.)

\section{Keeping terms $\mathcal{O}(T^2)$}
\label{OT^2}
The equations of motion now are, once again,
\begin{eqnarray}
\partial_u\partial_vW=-\partial_u T\partial_v T \\
\partial^2_uW-\partial_u\sigma_0\partial_uW=-(\partial_uT)^2\\
\partial^2_vW-\partial_u\sigma_0\partial_vW=-(\partial_vT)^2
\end{eqnarray}
and
\begin{equation}
\partial_u\partial_v[W-2\Phi]+4\partial_u\Phi\partial_v\Phi+\partial_uT\partial_vT+\frac{e^\sigma}{4}V(T)-\frac{2e^\sigma}{\alpha'}=0
\end{equation}
which may be rewritten, using $\partial_u\partial_vW=-\partial_u T\partial_v T$ as
\begin{equation}
-2\partial_u\partial_v\Phi+4\partial_u\Phi\partial_v\Phi+\frac{e^\sigma}{4}V(T)-\frac{2e^\sigma}{\alpha'}=0
\end{equation} 
where, as already mentioned, $W=\sigma_1-2\Phi_1$.

In addition, we have an equation of motion for the tachyon, which happens to be
\begin{equation}
-2\nabla^2 T+4\nabla\Phi\nabla T+V'(T)=0
\end{equation}
which we solve for the allowed form of $T$, as in subsection \ref{T}, to obtain, assuming separability, $T(u,v)=U(u)V(v)$,
\begin{equation}
T=\frac{\sqrt{A}}{\sqrt{(u-\kappa\sqrt{c_M})(v+\sqrt{c_M}/\kappa)}}
\end{equation}
where $A$ and $\kappa$ are constants independent of $u,v$ and $\kappa$ originates from the equation
\begin{eqnarray}
\frac{1+2u\frac{U'}{U}}{2\sqrt{c_M}\frac{U'}{U}}=-\frac{2\sqrt{c_M}\frac{V'}{V}}{1+2v\frac{V'}{V}}=\kappa ~.
\end{eqnarray} 

Later, we will find it convenient to rescale our coordinate system to $\bar{u}=u/\kappa\sqrt{c_M}$ and $\bar{v}=v\kappa/\sqrt{c_M}$ to get 
\begin{equation}
T=\frac{\sqrt{\bar{A}}}{\sqrt{(\bar{u}-1)(\bar{v}+1)}}
\end{equation}

It may be shown, using the form of $\sigma_0$ as derived in \cite{Mandal:1991tz}, that the `$W$' equations of motion reduce to 
\begin{eqnarray}
W_{uv}=-T_u T_v\\
W_{uu}+\frac{v}{uv+c_M} W_u=-T^2_u\\
W_{vv}+\frac{u}{uv+c_M} W_v=-T^2_v\\
1-(uv+c_M)\Phi_{1uv}+2(u\Phi_{1u}+v\Phi_{1v})+e^{\sigma_1}(\alpha' V(T)/2-1)=0
\end{eqnarray}
where $c_M=M\alpha'/2$.

For details of the solution of this system of equations, see Appendix \ref{W}.

We still have to solve for $W$ and $\Phi_1$ and, from these, $\sigma_1$.

Now, dropping the $O(T^4)$ terms in $W_u, W_v$ (see equations \ref{W_u}, \ref{W_v}), we are left with (in rescaled variables $u,v$)
\begin{eqnarray}
W_u=\frac{\beta v}{uv+1}\\
W_v=\frac{\beta u}{uv+1}
\end{eqnarray}
We can immediately guess that the solution has the form 
\begin{equation}
W(u,v)=\alpha+\beta\ln(uv+1)
\end{equation}
That is,
\begin{equation}
\sigma_1-2\Phi_1=\alpha+\beta\ln(uv+1)
\end{equation}
or
\begin{equation}
\sigma_1=2\Phi_1+\alpha+\beta\ln(uv+1)
\end{equation}
so
\begin{equation}
e^{\sigma_1}=e^\alpha e^{2\Phi_1}(uv+1)^\beta
\end{equation}

Here we can set $\alpha=0$ without affecting our results, as it only contributes an overall constant scale factor to the full metric, and further assume $\Phi_1$ to be small. Then we have
\begin{equation}
e^{\sigma_1}=(uv+1)^\beta (1+2\Phi_1)
\end{equation}
(We note that in Appendix \ref{W} we have demonstrated that $\beta$ must equal $-\frac{A}{4}$.)

Now we solve for $\Phi_1$ from the equation
\begin{equation}
1-(uv+c_M)\Phi_{1uv}+2(u\Phi_{1u}+v\Phi_{1v})+e^{\sigma_1}(\alpha' V(T)/2-1)=0
\end{equation}
In our rescaled variables, the equation becomes
\begin{equation}
1-(uv+1)\Phi_{1uv}+2(u\Phi_{1u}+v\Phi_{1v})-\frac{e^{\sigma_1}}{4}(T^2+2)=0
\end{equation}
or
\begin{equation}
2(uv+1)\Phi_{1uv}+(u\Phi_{1u}+v\Phi_{1v})=1-\frac{1}{2}\left[1+\frac{A}{(u-1)(v+1)}\right](uv+1)^\beta (1+2\Phi_1) ~\label{phi1}
\end{equation}
assuming $\Phi_1$ to be small and $\alpha=0$.

For details of the solution for of eqn (\ref{phi1}), $\Phi_1$, see Appendix \ref{Phi_1}.


\section{The perturbed metric: its near horizon behaviour}
\label{nearhorizonbehaviour}

The solution for $\sigma_1$ may be obtained via a solution of $\Phi_1$ (see Appendix \ref{Phi_1}) and the equation (\ref{eqnsigma1}).

The unperturbed metric was:
\begin{equation}
ds^2=e^{\sigma_0} du dv 
\end{equation}
where
\begin{eqnarray}
e^{\sigma_0} du dv= \frac{du dv}{\frac{2uv}{\alpha'}+M}=\frac{\alpha'}{2}\frac{d\bar{u}d\bar{v}}{\bar{u}\bar{v}+1} 
\end{eqnarray}
Then near the horizon, taking into account the metric perturbation $\sigma_1$, 
\begin{eqnarray}
ds^2=e^{\sigma_0+\sigma_1}du dv=\frac{\alpha'}{2}\frac{\exp\left[a_0\sum_{n=0}^\infty (\bar{u}\bar{v})^n/(n!)^2 \right]}{(1+\bar{u}\bar{v})^{1+A/4}} d\bar{u}d\bar{v} ~\label{sigma1}
\end{eqnarray}
where, we remind ourselves, the parameters $a_0$ and $A$ characterize the tachyon solution; $a_0$ is defined in the equation (\ref{def_a_0}). Setting these two to $0$ recovers the unperturbed metric. The horizon appears smooth as we show in the next subsection by evaluating the Kretschmann scalar.

\subsection{The nature of the horizon}
Using the fact that in 1+1 dimensions, the Weyl tensor vanishes, and the Riemann tensor has only one independent nonzero component, namely $R_{uvuv}=R_{vuvu}=-R_{uvvu}=-R_{vuuv}$, we compute this component to be
\begin{equation}
R_{uvuv}=-\frac{1}{3}e^\sigma\partial_u\partial_v\sigma
\end{equation}
and therefore 
\begin{equation}
R^{uvuv}=-\frac{4}{3}e^{-3\sigma}\partial_u\partial_v\sigma
\end{equation}

So the Kretschmann scalar
\begin{eqnarray}
\nonumber K=R_{abcd}R^{abcd}=\frac{64}{9}e^{-2\sigma}(\partial_u\partial_v\sigma)^2\\
=\left(\frac{2}{3} R\right)^2
\end{eqnarray}
where the Ricci scalar may be shown to be
\begin{equation}
R=4e^{-\sigma}\partial_u\partial_v\sigma
\end{equation}

It follows that the Kretschmann scalar will be non-singular wherever/whenever the Ricci scalar is. Now, from equations (\ref{appbsigma1}) and (\ref{appbsigmah}), it is obvious that {\it on} the horizon, the Ricci scalar $R = const$. Consequently, the horizon is non-singular under ${\cal O}(T^2)$ perturbations of the equations of motion for the black hole. It also follows from these equations that spacetime in the vicinity of the horizon is also non-singular, the only singularity is that which ensues for the $T=0$ situation.

\section{Conclusions}

In this paper, we have considered the extension of the 1+1 dimensional black hole in zero tachyon background \cite{Witten:1991yr, Mandal:1991tz, Rocek:1991vk} to the case where tachyon field as well as the tachyon potential are nontrivial, but perturbatively small, such that the conformal factor of the metric and the dilaton fields also change perturbatively linearly.

If the metric and dilaton perturbations are labeled $\sigma_1$ and $\Phi_1$, we show that $\sigma_1=2\Phi_1+\beta\ln(uv+1)$, where  $\beta=-A/4$ is a constant characterizing the tachyon and $u,v$ are Kruskal-Szekeres-like coordinates, as used, for example, in \cite{Mandal:1991tz}. 

We also depart from the standard ansatz for the tachyon adopted following references \cite{Witten:1991yr, Mandal:1991tz}. Assuming that the tachyon $T=T(u,v)$ is a separable function of the coordinates $u,v$, we evaluate the form of the tachyon hair that is permissible and the form of the perturbed metric. 

From this we evaluate the Kretschmann scalar for the black hole, and find it to be non-singular at the horizon which, incidentally, coincides with the horizon in the unperturbed case. The tachyon perturbation solution and the tachyon potential have been determined self-consistently to the order of our approximations. So long as the tachyon field remains perturbative, the black hole geometry still appears to be meaningful, and therefore the issue of an exact, `perturbed' conformal field theory, as the quantum description (`ultraviolet completion') of the classical black hole geometry, remains alive.   


\appendix

\section{Solution of the Set of Equations for $W=\sigma_1-2\Phi_1$}
\label{W}

Using the form of $\sigma_0$ as derived in \cite{Mandal:1991tz},  the equations of motion involving $W=\sigma_1-2\Phi_1$ reduce to 
\begin{eqnarray}
W_{uv}=-T_u T_v\\
W_{uu}+\frac{v}{uv+c_M} W_u=-T^2_u\\
W_{vv}+\frac{u}{uv+c_M} W_v=-T^2_v\\
1-(uv+c_M)\Phi_{1uv}+2(u\Phi_{1u}+v\Phi_{1v})+e^{\sigma_1}(\alpha' V(T)/2-1)=0
\end{eqnarray}
where $c_M=M\alpha'/2$.

Our strategy in solving for $W=W(u,v)$ is outlined as follows: 

\begin{itemize}
\item We have no reason to assume that $W$ is a separable function of $u,v$. 
\item The two equations involving $W_{uu}, W_{vv}$ are really first order equations in $W_u, W_v$. We can solve these separately for $W$ and match solutions after the integrations are done. 
\item To check for consistency with the equation $W_{uv}=-T_u T_v$, we can differentiate $W_u$ with respect to $v$ and $W_v$ with respect to $u$, and check for equality, assuming the space is simply connected.
\end{itemize}

In the rescaled coordinates $\bar{u},\bar{v}$, we have the following relations:
\begin{eqnarray}
\bar{A}=A/c_M, \hspace{1cm} uv+c_M=c_M(\bar{u}\bar{v}+1)\\
\partial_u=\frac{1}{\kappa\sqrt{c_M}}\partial_{\bar{u}}, \hspace{1cm} \partial_v=\frac{\kappa}{\sqrt{c_M}}\partial_{\bar{v}}\\
u\partial_u=\bar{u}\partial_{\bar{u}} \hspace{1cm} v\partial_v=\bar{v}\partial_{\bar{v}} \hspace{1cm} \partial_u\partial_v=\frac{1}{c_M}\partial_{\bar{u}}\partial_{\bar{v}}\\
(uv+c_M)\partial_u\partial_v=(\bar{u}\bar{v}+1)\partial_{\bar{u}}\partial_{\bar{v}}
\end{eqnarray} 

Then the $W$ equations will be seen to retain the same form as before, so for simplicity we will drop the bars on $u,v$ and $A$, reinserting them only as and when necessary. Then from
\begin{equation}
T=\frac{\sqrt{A}}{\sqrt{(u-1)(v+1)}}
\end{equation}
we can show that
\begin{eqnarray}
T^2_u=A/4 (u-1)^{-3}(v+1)^{-1}\\
T^2_v=A/4 (u-1)^{-1}(v+1)^{-3}\\
T_u T_v=T^2 /4
\end{eqnarray}

Defining $X=W_u, Y=W_v$,  
\begin{eqnarray}
X_u+\frac{v}{uv+1}X=-A/4(u-1)^{-3}(v+1)^{-1}\\
Y_v+\frac{u}{uv+1}Y=-A/4(u-1)^{-1}(v+1)^{-3}
\end{eqnarray}
Both these first order equations may be shown to have an integrating factor of $(uv+1)$. Then, we have:
\begin{eqnarray}
\partial_u[(uv+1)X]=-\frac{A}{4}\frac{uv+1}{(u-1)^3 (v+1)}\\
\partial_v[(uv+1)Y]=-\frac{A}{4}\frac{uv+1}{(u-1) (v+1)^3}
\end{eqnarray}

Taking the first of these two equations:
\begin{equation}
\partial_u[(uv+1)X]=-\frac{A}{4}\frac{uv+1}{(u-1)^3 (v+1)}
\end{equation}
we can rewrite the right hand side to get
\begin{eqnarray}
\partial_u[(uv+1)X]=-\frac{A}{4}\frac{uv+1}{(u-1)^3 (v+1)}\\
=-\frac{A}{4(v+1)}\left[\frac{1}{(u-1)^3}+v\frac{u}{(u-1)^3}\right]\\
=-\frac{A}{4(v+1)}\left[\frac{1}{(u-1)^3}+v\frac{1}{(u-1)^2}+v\frac{1}{(u-1)^3}\right]\\
=-\frac{A}{4}\left[\frac{v}{v+1}\frac{1}{(u-1)^2}+\frac{1}{(u-1)^3}\right]
\end{eqnarray}
That is,
\begin{equation}
\partial_u[(uv+1)X]=-\frac{A}{4}\left[\frac{v}{v+1}\frac{1}{(u-1)^2}+\frac{1}{(u-1)^3}\right]
\end{equation}
Integrating with respect to $u$,
\begin{equation}
(uv+1)X=A_1(v)+\frac{A}{4}\frac{v}{(u-1)(v+1)}+\frac{A}{8(u-1)^2}
\end{equation}
where $A_1(v)$ is purely a function of $v$ to be determined later by matching solutions for $W$. Thus, since $X=W_u$, we have
\begin{equation}
\label{W_u}
W_u=\frac{A_1(v)}{uv+1}+\frac{A}{4}\frac{v}{(uv+1)(u-1)(v+1)}+\frac{A}{8(uv+1)(u-1)^2}
\end{equation}
Similarly, we may show that, after taking care of the signs correctly,
\begin{equation}
\label{W_v}
W_v=\frac{B_1(u)}{uv+1}+\frac{A}{4}\frac{u}{(uv+1)(u-1)(v+1)}-\frac{A}{8(uv+1)(v+1)^2}
\end{equation}
where $B_1(u)$ likewise is a function of $u$ alone, to be determined by matching solutions later. Now, before we integrate equations (\ref{W_u}, \ref{W_v}), which looks a tedious process, an alternative is to set up a constraint equation by differentiating these equation by demanding consistency with $W_{uv}=-T_u T_v$. Let us see what we get.
\begin{eqnarray}
{(W_u)}_v=\frac{{A'}_1(v)}{uv+1}-\frac{A_1(v)u}{(uv+1)^2}+\frac{A(1-uv^2)}{4(u-1)(v+1)^2(uv+1)^2}-\frac{Au}{8(u-1)^2(uv+1)^2}\\
{(W_v)}_u=\frac{{B'}_1(u)}{uv+1}-\frac{B_1(u)v}{(uv+1)^2}-\frac{A(1+u^2v)}{4(u-1)^2(v+1)(uv+1)^2}+\frac{Av}{8(v+1)^2(uv+1)^2}
\end{eqnarray}
If these are to match, we must at least have $A_1(v)=\beta v, B_1(u)=\beta u$, where $\beta$ is a constant independent of $u,v$. This makes our equations look like this:
\begin{eqnarray}
{(W_u)}_v=\frac{\beta}{uv+1}-\frac{\beta uv}{(uv+1)^2}+\frac{A(1-uv^2)}{4(u-1)(v+1)^2(uv+1)^2}-\frac{Au}{8(u-1)^2(uv+1)^2}\\
{(W_v)}_u=\frac{\beta}{uv+1}-\frac{\beta uv}{(uv+1)^2}-\frac{A(1+u^2v)}{4(u-1)^2(v+1)(uv+1)^2}+\frac{Av}{8(v+1)^2(uv+1)^2}
\end{eqnarray}
It seems harder to match the last two terms, but we must recall that we have dropped terms $O(T^3)$. Perhaps that has some relevance in the present situation.

Then it turns out that
\begin{equation}
W_{uv}-W_{vu}=\frac{A(u+v)}{8(u-1)^2(v+1)^2(uv+1)}
\end{equation}
Recalling that
\begin{equation}
T=\frac{\sqrt{A}}{\sqrt{(u-1)(v+1)}}
\end{equation}
we have
\begin{equation}
W_{uv}-W_{vu}=\frac{T^4 (u+v)}{8A(uv+1)}
\end{equation}
aside from a few factors of $c_M$ which arise from the rescaling of coordinates, but this is a minor issue here. Now $(u+v)/(uv+1)$ is $O(T)$ if we consider the solution for $T$, this term is like $O(T^5)$. While this looks like some evidence, however minute, for torsion in this space, if we ignore terms $O(T^3)$ as intended, we have left (assuming these are still not negligible) 
\begin{eqnarray}
W_{uv}=W_{vu}=\frac{\beta}{uv+1}-\frac{\beta uv}{(uv+1)^2}=\frac{\beta}{(uv+1)^2}
\end{eqnarray}
We still have to check if this matches with 
\begin{eqnarray}
-T_u T_v=-\frac{A/4}{(uv+u-v-1)^2}=-\frac{A/4}{(u-1)^2 (v+1)^2} 
\end{eqnarray}
For large $u,v$ if these are to agree, we can fix the constant $\beta=-A/4$, so that one possibility is that
\begin{eqnarray}
A_1(v)=-Av/4, \hspace{1cm} B_1(u)=-Au/4
\end{eqnarray}
but $A$ is still an arbitrary constant of integration. The matching may look a bit forced, but we observe that $T_u T_v$ is $O(T^4)$ and is therefore negligible in our scheme of things (where we retain terms upto $O(T^2)$). 

Now, dropping the $O(T^4)$ terms in $W_u, W_v$, we are left with
\begin{eqnarray}
\label{W_u}
W_u=\frac{\beta v}{uv+1}\\
\label{W_v}
W_v=\frac{\beta u}{uv+1}
\end{eqnarray}
We can immediately guess that the solution has the form 
\begin{equation}
W(u,v)=\alpha+\beta\ln(uv+1)
\end{equation}

Considering the fact that $e^\alpha$ only contributes an overall constant factor to the full metric, we can set $\alpha=0$, and since we have already shown that $\beta=-\frac{A}{4}$, we have
\begin{equation}
\label{eqnsigma1}
\sigma_1=2\Phi_1-\frac{A}{4}\ln(uv+1)
\end{equation}
in our rescaled variables.

\section{Solution for $\Phi_1$, $\sigma_1$}
\label{Phi_1}

The differential equation satisfied by $\Phi_1$ is
\begin{equation}
2(uv+1)\Phi_{1uv}-2(u\partial_u+\partial_v)\Phi_1-(1+uv)^{-A/4}(1+T^2/4)(1+2\Phi_1)=-1
\end{equation}

Defining $1+2\Phi_1=y(u,v)$,
\begin{equation}
(uv+1)y_{uv}-(u\partial_u+\partial_v)y-(1+uv)^{-A/4}(1+T^2/4)y=-1
\end{equation}

There are two terms here that depend on $T^2$: one is $\left(1+T^2/4\right)$ and the other is $A$ (which is proportional to $T^2$. If we may assume that $T^2/4$ is negligible compared to $1$, but $A uv/4$, although small, is not, then our equation becomes
\begin{equation}
(uv+1)y_{uv}-(u\partial_u+v\partial_v)y-(1+\beta uv)y=-1
\end{equation}
where $\beta=-A/4$. Defining $uv=x$, we end up with the equation
\begin{equation}
(x+1)(\partial_x+x\partial_x^2)y-2x\partial_x y-(1+\beta x)y=-1
\end{equation}
or
\begin{equation}
x(x+1)y''+(1-x)y'-(1+\beta x)y=-1
\end{equation}

Here, dropping the source term, we have the homogeneous differential equation:
\begin{equation}
x(x+1)y''+(1-x)y'-(1+\beta x)y=0
\end{equation}
which we can recast as 
\begin{equation}
y''+\frac{1-x}{x(1+x)}y'-\frac{1+\beta x}{x(1+x)}y=0
\end{equation}
which has two regular singular points: $x=0$ ($uv=0$, at the horizon) and $x=-1$ ($uv=-1$ in rescaled coordinates, that is, at the singularity). Close to the horizon ($x=uv=0$), this last differential equation has the following structure:
\begin{equation}
y''+\frac{1}{x}y'-\frac{1}{x}y=0
\end{equation}
for the homogeneous equation, and
\begin{equation}
y''+\frac{1}{x}y'-\frac{1}{x}y=-\frac{1}{x}
\end{equation}
for the inhomogeneous equation.

These can be rewritten as
\begin{equation}
x y''+y'-y=0
\end{equation}
for the homogeneous equation, and
\begin{equation}
x y''+y'-y=-1
\end{equation}
for the inhomogeneous equation. 

The homogeneous equation may be solved by Frobenius' method to obtain a solution for the form
\begin{equation}
\label{def_a_0}
y_1(x)=a_0\sum_{n=0}^\infty\frac{x^n}{(n!)^2}
\end{equation}
and a second solution that converges wherever the first one does:
\begin{equation}
y_2(x)=y_1(x)\ln{x}+\sum_{n=0}^\infty A_n^*x^n
\end{equation}

The inhomogeneous equation near the horizon can be rewritten as
\begin{equation}
x^2 y''+x y'-x y=-x
\end{equation}
or, using the fact that $y=1+2\Phi_1$,
\begin{equation}
x^2\Phi_1''(x)+x\Phi_1'(x)-x\Phi_1(x)=0
\end{equation}
Cancelling a factor of $x$, we have 
\begin{equation}
x\Phi_1''(x)+\Phi_1'(x)-\Phi_1(x)=0
\end{equation}
This is something we have already solved, for the homogeneous equation. Using the equation (\ref{eqnsigma1}), we have 
\begin{equation}
\sigma_1=2\Phi_1-\frac{A}{4}\ln{(uv+1)}
\end{equation}
in rescaled variables, we must have $\sigma_1$ of the form 
\begin{equation}
\sigma_1=\beta\ln{(1+uv)}+a_0 \sum_{n=0}^\infty\frac{(uv)^n}{(n!)^2} ~\label{appbsigma1}
\end{equation}
Close to the horizon, then,
\begin{equation}
\sigma_1\approx \beta uv+a_0+a_0 uv ~\label{appbsigmah}
\end{equation}
or
\begin{equation}
\label{nearhorizon_sigma_1}
\sigma_1\approx a_0+(a_0+\beta) uv
\end{equation}

Using these expansions, it is possible to show that 
the Ricci scalar, given by
\begin{equation}
R=4e^{-\sigma}\partial_u\partial_v\sigma
\end{equation}
is finite at the horizon, and so is the Kretschmann scalar
\begin{eqnarray}
\nonumber K=R_{abcd}R^{abcd}=\frac{64}{9}e^{-2\sigma}(\partial_u\partial_v\sigma)^2\\
=\left(\frac{2}{3} R\right)^2
\end{eqnarray}

\section*{Acknowledgments}
We would like to thank Dr Saugata Bhattacharyya for occasional discussions. One of us (AB) would like to thank the physics departments of Charuchandra College, Rammohan College, Ramakrishna Mission Residential College (Narendrapur) and Ramakrishna Mission Vivekananda University for hospitality during the course of this work.


\end{document}